\documentclass[preprint2,twoside]{hwo}

\usepackage{amsmath,amssymb}
\usepackage{graphicx}
\usepackage{CJK}
\usepackage{lipsum}

\input{hwo.h}

\begin{document}

\title{\textbf{\LARGE White dwarfs as probes of extrasolar planet compositions and fundamental astrophysics}}
\author {\textbf{\large Siyi Xu \begin{CJK*}{UTF8}{gbsn}(许\CJKfamily{bsmi}偲\CJKfamily{gbsn}艺\end{CJK*})~$^{1}$, Martin Barstow~$^2$, Andy Buchan~$^3$, \'{E}rika Le Bourdais~$^{4,5}$ and~Patrick~Dufour~$^{4,5}$ }}
\affil{$^1$\small\it Gemini Observatory/NOIRLab, 950 N Cherry Ave, Tucson, AZ 85719}
\affil{$^2$\small\it Department of Physics and Astronomy, University of Leicester, University Road, Leicester, LE1 7RH, UK;}
\affil{$^3$\small\it Department of Physics, University of Warwick, Coventry CV4 7AL, UK}
\affil{$^4$\small\it Institut Trottier de recherche sur les Exoplan\`etes (IREx), Universit\'e de Montr\'eal, Montr\'eal, QC H3C 3J7, Canada}
\affil{$^5$\small\it D\'epartement de physique, Universit\'e de Montr\'eal, Montr\'eal, QC H3C 3J7, Canada}

\author{\footnotesize{\bf Endorsed by:}
Nicholas	Ballering	(Space Science Institute), 
Solen	Balman	(Istanbul University), 
Juliette	Becker	(UW-Madison), 
Simon	Blouin	(University of Victoria), 
Howard	Bond	(Penn State University), 
Amy	Bonsor	(University of Cambridge), 
Jean-Claude	Bouret	(Laboratoire d'Astrophysique de Marseille), 
Kara	Brugman	(Arizona State University), 
Roberto	Capuzzo Dolcetta	(University of Roma), 
Aarynn	Carter	(STScI), 
Sarah	Casewell	(University of Leicester), 
Luke	Chamandy	(National Institute of Science Education and Research), 
Melanie	Crowson	(American Public University), 
Nathalie	Degenaar	(University of Amsterdam), 
Vikram	Dhillon	(University of Sheffield), 
Caroline	Dorn	(ETH Z\"urich), 
Trent	Dupuy	(University of Edinburgh), 
Jay	Farihi	(University College London), 
Luca	Fossati	(Space Research Institute, Austrian Academy of Sciences), 
Emma 	Friedman	(NASA GSFC), 
Boris	G\"{a}nsicke	(University of Warwick), 
Hongwei	Ge	(Yunnan Observatories, Chinese academy of sciences), 
Stephan	Geier	(University of Potsdam), 
Christopher	Glein	(Southwest Research Institute), 
Diego	Godoy-Rivera	(Instituto de Astrof\'{i}sica de Canarias), 
Joseph	Guidry	(Boston University), 
Na'ama	Hallakoun	(Weizmann Institute of Science), 
Zhanwen	Han	(Yunnan Observatories, Chinese academy of sciences), 
Natalie	Hinkel	(Louisiana State University), 
Steve	Howell	(NASA), 
Fran	Jim\'{e}nez-Esteban	(CAB, CSIC-INTA), 
David	Jones	(Instituto de Astrof\'{i}sica de Canarias), 
Steven	Kawaler	(Iowa State University), 
Cenk	Kayhan	(Kayseri University), 
Souza Oliveira	Kepler	(Universidade Federal do Rio Grande do SUl, Brazil), 
Mukremin	Kilic	(University of Oklahoma), 
Agnes	Kim	(Penn State University), 
Ji\v{r}\'{i}	Krti\v{c}ka	(Masaryk University), 
Eunjeong	Lee	(EisKosmos CROASAEN, Inc.), 
Uri	Malamud	(Technion Israel Institute of Technology), 
Elena	Manjavacas	(STScI), 
Carl	Melis	(University of California, San Diego), 
Jaroslav	Merc	(Astronomical Institute of Charles University), 
Alexander	Mustill	(Lund University, Sweden), 
Faraz	Nasir Saleem	(Egypt Space Agency, EgSA), 
Gijs	Nelemans	(Radboud University), 
Terry	Oswalt	(Embry-Riddle Aeronautical University), 
Quentin	Parker	(The University of Hong Kong), 
Steven	Parsons	(University of Sheffield), 
Alberto	Rebassa-Mansergas	(Universitat Polit\`{e}cnica de Catalunya), 
Seth	Redfield	(Wesleyan University), 
Claudia	Rodrigues	(National Institute for Space Research, Brazil), 
Laura	Rogers	(NOIRLab), 
Farid	Salama	(NASA Ames Research Center), 
Matthias R. 	Schreiber	(Universidad Tecnica Federico Santa Maria), 
Roberto	Silvotti	(INAF-Osservatorio Astrofisico di Torino), 
J. Allyn	Smith	(Austin Peay State University), 
Melinda	Soares-Furtado	(UW-Madison), 
Frank	Soboczenski	(University of York \& King's College London), 
Jie	Su	(Yunnan Observatories, Chinese Academy of Sciences), 
Andrew	Swan	(University of Warwick), 
Paula	Szkody	(U of Washington), 
Jes\'{u}s A.	Toal\'{a}	(IRyA - UNAM), 
Silvia	Toonen	(University of Amsterdam), 
Dimitri	Veras	(University of Warwick), 
Haiyang	Wang	(University of Copenhagen), 
Austin	Ware	(Arizona State University), 
Klaus	Werner	(Universit\"{a}t T\"{u}bingen), 
Peter	Wheatley	(University of Warwick), 
David	Wilson	(University of Colorado)
}



\begin{abstract}
{White dwarfs represent the most common end stage of stellar evolution and are important for a range of astrophysical questions. The high-resolution ultraviolet spectroscopic capability of the {\it Habitable World Observatory} (HWO) offers a unique capability to characterize white dwarfs. In this documents, we focus on two specific science cases for HWO -- white dwarfs as probes of extrasolar planet compositions, and fundamental astrophysics. HWO will have the sensitivity to measure a suite of heavy elements, such as S, C, O, Fe, and Si, in a large sample of polluted white dwarfs to constrain the water content and the light elements in the cores of extrasolar planets. HWO can also be used to search for any small variation on the fine structure constant in the presence of strong gravity. Both science cases require a minimum resolving power of 60,000, and a ultraviolet coverage down to at least 900 Angstrom.
  }
  \\
  \\
\end{abstract}

\vspace{2cm}

\section{Science Goal}

\subsection{White dwarfs as probes of extrasolar planet compositions}

It is now widely accepted that many white dwarfs are actively accreting planetary material from exo-asteroids, exo-comets, and occasionally exo-planets. These white dwarfs are so called polluted white dwarfs -- exoplanetary material is polluting the white dwarf's pure hydrogen or helium atmosphere. Hence, spectroscopic observations of polluted white dwarf atmospheres can reveal the incidence of planetary systems and, uniquely, the bulk composition of extra-solar planetary material \citep[e.g.][]{JuraYoung2014, Swan2019, Xu2024b}. Extracting this information from the observations depends on a thorough understanding of the evolution of white dwarfs, including their photospheric compositions and the physical mechanisms that determine these. This is a unique tool because bulk compositions of solar system objects, such as Mars, Venus, asteroids, and even Earth, are hard to measure, not to mention extrasolar bodies. 

There is also the puzzle of the circumstellar material seen in the UV spectra of many hot white dwarfs \citep[e.g.][]{Lallement2011, Dickinson2012, Gaensicke2012}. It is possible that this circumstellar material may also be associated with closely orbiting planetary debris, which would be sublimated around hot stars. Alternatively, it could be remnants from earlier phases of mass-loss, part of the cosmic recycling of processed material, enriching the local interstellar medium (LISM). Radiation from hot white dwarfs also affects the LISM ionisation balance surrounding them. Some high ionisation lines detected in white dwarf spectra are interstellar rather than circumstellar, tracing the physical state of the LISM and testing models of the formation and evolution of the Local Bubble.

\begin{figure*}[h]
\begin{center}
\includegraphics[width=0.85\textwidth,trim={0 2cm 0 2cm},clip]{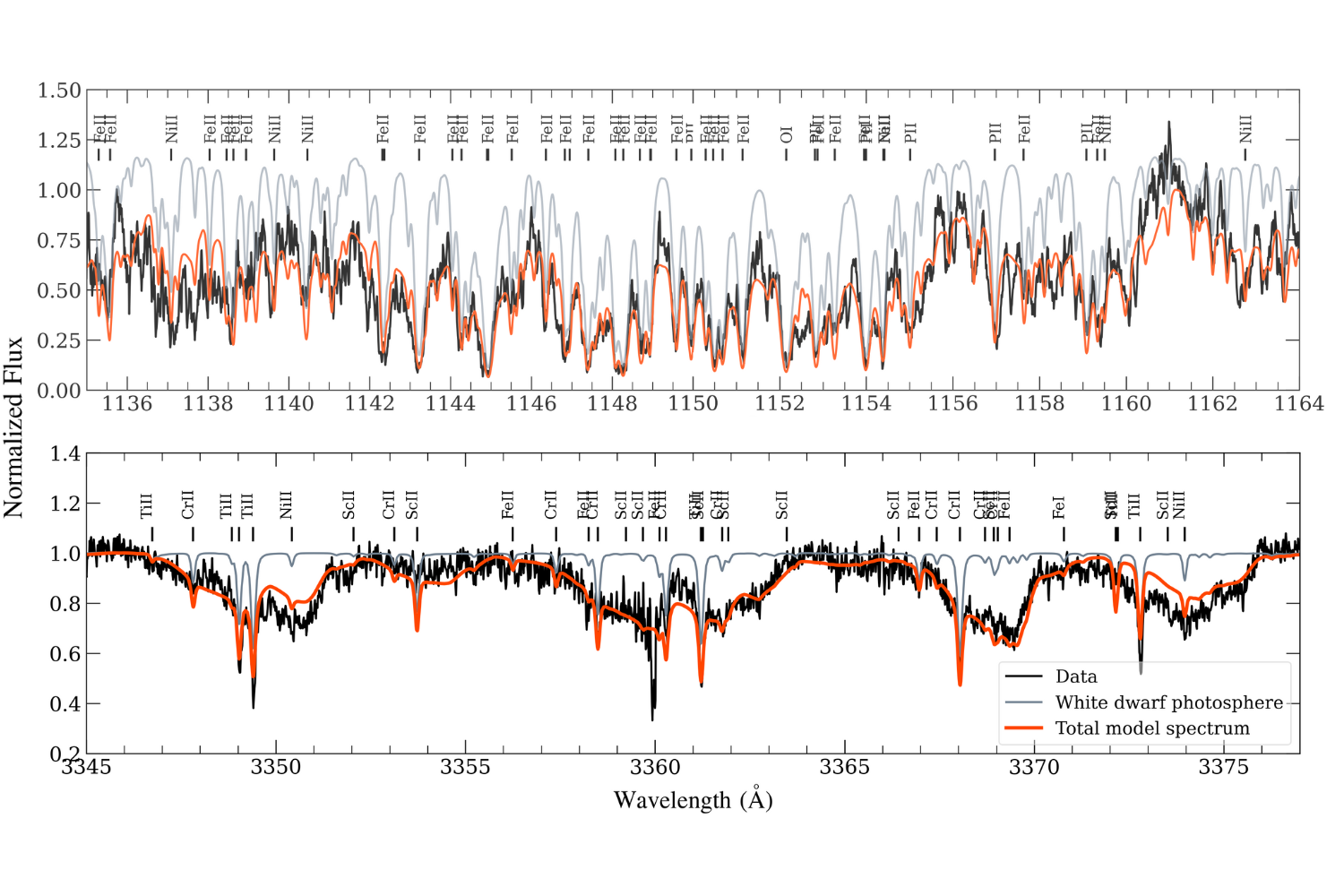}
\caption{\small HST/COS (upper panel) and Keck/HIRES (lower panel) spectra of WD~1145+017, a heavily polluted white dwarf that also has circumstellar material \citep{LeBourdais2024}. The black line is the data; the grey and red line represents a model with white dwarf and white dwarf with circumstellar, respectively. The density of spectral lines is much higher in the ultraviolet, making it essential for detecting minor and trace elements in polluted white dwarfs.
\label{fig1}
}
\end{center}
\end{figure*}

For white dwarfs hotter than 10,000K where there is sufficient flux from the white dwarf at the ultraviolet wavelengths, spectral lines in the ultraviolet tend to be stronger and more abundant compared to the optical. However, ultraviolet radiation is absorbed from Earth's atmosphere, and is therefore only observable from space. Figure~\ref{fig1} shows the spectrum of a heavily polluted white dwarf WD~1145+017; most of the white dwarf photospheric lines are below 4000 Angstrom. The Cosmic Origins Spectrograph (COS) on the Hubble Space Telescope (HST) is the only instrument available right now with enough sensitivity and spectral resolution to study polluted white dwarfs in the ultraviolet. The Habitable Worlds Observatory (HWO) will be able to revolutionize this field by observing a large number of polluted white dwarfs and measuring the composition of extrasolar bodies. Additionally, the strongest lines for the volatile elements (C, N, O) are all at ultraviolet wavelengths, making these ultraviolet observations crucial for questions regarding volatile loss and habitability in exoplanetary systems \citep{Rogers2024a}.

\subsection{Does Physics change in the presence of strong gravity?}

General Relativity (GR) has passed all weak-field observational and experimental tests to date. Nevertheless, the theory predicts singularities on small scales and is incompatible with quantum field theory. To be compatible with cosmological observations, the theory also requires most of the energy content of the Universe to be in the form of an unknown dark energy. These things suggest that GR may ultimately become the low-energy limit of some future more fundamental quantum gravity unification theory. Several such theories allow the possibility of space and time variations in the low energy 'constants` of Nature, either because of the presence of extra space dimensions or the non-uniqueness of the quantum vacuum state for the universe. Searches for departures from standard GR, particularly in stronger-field situations, are important to test such ideas. Provided objects are not too relativistic, the total mass and total scalar charge are just proportional to the number of nucleons in the object. This is the case for stars and planets. Near massive gravitating bodies, different types of couplings between scalar fields and other fields can lead to an increase or decrease in the coupling constant strengths \citep{Magueijo2002}. For small variations, the scalar field variation is proportional to the change in the dimensionless gravitational potential, $GM/Rc^2$ (G is the gravitational constant, M the mass of the object, R its radius and c the speed of light) and hence the compactness (M/R) of an object. Compact objects with high mass and small radius, could exhibit detectable variations of fundamental constants \citep[e.g.][]{FlambaumShuryak2008}.

White dwarfs are an important group of objects with high surface gravitational potential where constant variations might be detected. When a stellar mass is compressed into a radius comparable to Earth's, strong gravitational fields are generated, and observations
allow us to search for new physics in the presence of gravitational accelerations $GM/Rc^2$ $\approx$ 10$^4$ times stronger than on Earth. Their spectra can contain many narrow absorption lines for which precisely observed wavelengths can be measured and compared to expected values.  While neutron stars have much higher gravitational fields (log $\approx$ 14.0 - 15.0) than white dwarfs (log $\approx$ 7.0 - 9.0), gravitational broadening smooths the lines into invisibility. Therefore, from a practical perspective, only white dwarfs can be used to search for possible variations in the fundamental constants.

\citet{Berengut2013} placed constraints on the fractional variation of the fine structure constant $\alpha$, $\Delta \alpha / \alpha$, using \ion{Fe}{5} and \ion{Ni}{5} lines in white dwarf spectra, in a dimensionless gravitational potential a factor of $\approx$ 5 x 10$^5$ stronger than terrestrial. Further measurements of $\Delta \alpha / \alpha$ have been reported in \citet{Bainbridge2017}.

\cite{Xu2013b} discovered the presence of numerous Werner and Lyman band H$_2$ lines in $\approx$ 11,000 K white dwarf spectra. \citet{Bagdonaite2014} exploited this to make measurements of the proton to electron mass ratio in strong gravity. \citet{Bagdonaite2014}, compared the wavelengths of Lyman lines (the weaker Werner lines were not used) with accurate laboratory measurements. Because different Lyman transitions have different sensitivities to changes in the proton to electron mass ratio, $\mu$ = m$_p$/m$_e$, where m$_p$ and m$_e$ are the proton and electron mass, respectively, any change in $\mu$ would generate a unique pattern unlike a simple gravitational redshift or stellar motion effect.

\subsection{Overarching science questions}
\begin{itemize}
    \item What are the abundances of elements in the atmospheres of white dwarfs?
    \item How do the abundances of elements in the atmospheres of white dwarfs relate to general properties of the system?
    \item What does the material white dwarfs accrete tell us about their current and past systems of planets, asteroids, comets, and debris?
    \item What can we learn about their exoplanet family?
    \item Are the values of fundamental physical constants affected by strong gravity?
\end{itemize}

Related 'big board` Astro2020 topics
\begin{itemize}
    \item What are the Demographics of Planets Beyond the Reach of Current Surveys?
    \item What are the properties of individual planets, and which processes lead to planetary diversity?
    \item Fundamental physics
    \item What are the most extreme stars and stellar populations?
    \item 'Industrial Scale` Spectroscopy
\end{itemize}

\subsection{The need for Habitable Worlds Observatory}

Access to the ultraviolet wavelength range is crucial for both projects. The transitions of many important atomic, ionic, and molecular species are only seen in this wavelength range. This coverage needs to be coupled with high signal-to-noise, requiring an aperture much larger than that of current ultraviolet missions such as HST.

\section{Science Objective}

\subsection{Extrasolar planet composition objective \label{sec2:planet}}

Studying polluted white dwarfs can directly measure the chemical composition of extrasolar bodies, which is a fundamental property that directly affects its habitability and is closely linked to the main goal of HWO. Equipped with this powerful tool, we can use HWO to obtain a statistically significant sample of extrasolar planetary compositions. There are many things to explore, and here we focus on two key questions:

\begin{enumerate}
    \item \textbf{How much water do extrasolar planets have?} Water is crucial to all life on Earth, and the presence of water in an extrasolar planet has strong implications on the planet's habitability. While the James Webb Space Telescope might be able to detect water in a planet's atmosphere \citep{Moran2023}, measuring the total water fraction is difficult due to the degeneracies in planet models \citep{DornLichtenberg2021}. Polluted white dwarfs offer an alternative way to assess the water content in extrasolar planets.
    \item \textbf{What are the light elements in the cores of extrasolar planets?} Earth is differentiated into a core, mantle, and crust. It has long been suggested that there are some `light' elements, such as sulfur, silicon, oxygen, carbon and hydrogen, in Earth's core but it is difficult to know the exact composition \citep{Hirose2021}. If the typical light element content of extrasolar cores could be constrained, we may in turn constrain the likely formation pathway of Earth and other Solar System bodies. Differentiation and core formation appears to be common in extrasolar planetary bodies from polluted white dwarfs studies \citep{Bonsor2020,Doyle2020}. By having a large sample of polluted white dwarfs, we hope to determine the amount of light elements in extrasolar planetary cores. 
\end{enumerate}

\textbf{Observational Objective:} Currently, HST/COS is the best instrument for polluted white dwarf studies. There are only 15 polluted white dwarfs with detections of S, C, O, Mg, Fe, and Si, as shown in Table~\ref{tab:number_PWD} \citep{Williams2024}. In order to answer the science questions above, progress is needed both in terms of having higher quality data and better white dwarf modeling to decrease the abundance uncertainty. In order to make major progress, we need a sample of at least 45 polluted white dwarfs with detections of S, C, O, Mg, Fe, and Si and an abundance uncertainty of each element to be within 0.1 dex.
\begin{table*}[ht]
\caption{Number of polluted white dwarfs for extrasolar planet composition studies, see Sections \ref{sec2:planet} and \ref{sec3:planet} for details.}
    \label{tab:number_PWD}
\smallskip
\begin{center}
{\small
\begin{tabular}{lccc}  
\tableline
\noalign{\smallskip}
Physical Parameter & State of the Art & Substantial Progress & Major Progress\\
  & &(Enabling)& (Breakthrough)\\
\noalign{\smallskip}
\tableline
\noalign{\smallskip}
Number of polluted WDs with measurements of S, C, O, Fe and Si & 15 & 45 & 45 \\
        \noalign{\smallskip}
\tableline
\noalign{\smallskip}
Abundance uncertainty (both from data and model) & 0.3 dex & 0.2 dex & 0.1 dex\\

\noalign{\smallskip}
\tableline\
\end{tabular}
}
\end{center}
\end{table*}

\subsection{Fundamental physics objective}
Some theories of gravity predict that the well-known fundamental physical constants might have different values according to the strength of the local gravitational field. The values of two of these, the fine structure constant ($\alpha$) and the proton-electron mass ratio ($\mu$), can, in principle, be measured by comparing the observed wavelength of an atomic or molecular line with the value predicted by the basic physics of line formation. The main objective is to observe a sample of white dwarf stars covering the largest possible range of surface gravities to search for evidence of departures in the observed wavelengths of spectral lines from theoretical prediction. The results will be used to study how the fundamental physical constants $\alpha$ and $\mu$ vary in strong gravity, measuring real changes or placing limits on this phenomenon. 

\textbf{Observational Objective:}  Measure the wavelengths of high atomic mass and high ionization or molecular spectral lines for comparison with reference data from laboratory measurements and computational predictions.

\begin{figure*}[h]
    \centering
    \includegraphics[width=0.8\linewidth]{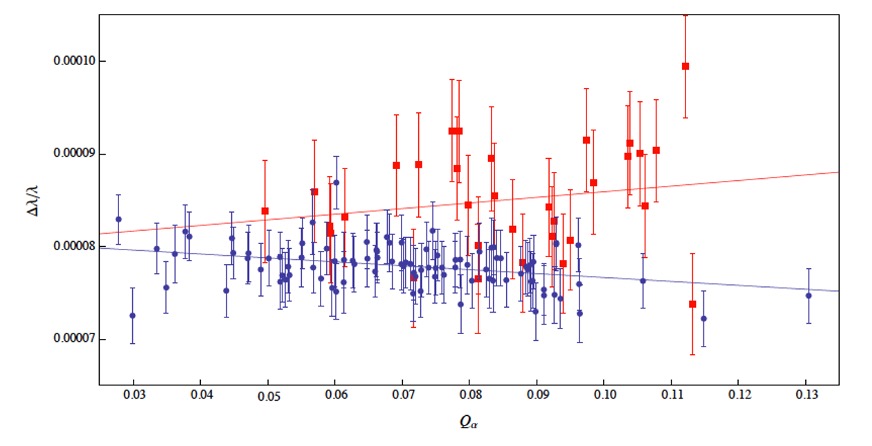}
    \caption{$\Delta \lambda$/$\lambda$  vs the sensitivity parameter Q$_\alpha$ for transitions in \ion{Fe}{5} (blue circles) and \ion{Ni}{5} (red squares). The slope of the lines gives $\Delta\alpha/\alpha=  4.2 \pm 1.6 \times 10^{-5}$ and $6.1 \pm 5.8 \times 10^{-5}$ for \ion{Fe}{5} and \ion{Ni}{5}, respectively. The slope seen in the \ion{Ni}{5} spectra is likely due to systematics present in the laboratory wavelength measurements rather than indicating a gravitational dependence of $\alpha$ \citep{Berengut2013}.}
    \label{fig: Qalpha}
\end{figure*}

The observational consequences of and changes in $\alpha$ and/or $\mu$ are predicted to be small shifts in the observed wavelengths of features in observed spectra. White dwarfs span a range of gravities covering approximately two orders of magnitude ($\log g = 7.0-9.0$). For $\alpha$, the shifts are largest and, potentially, easiest to detect for high atomic mass and high ionization species, e.g. for the Fe V/VI and Ni V/VI. The fractional wavelength shift for each line ($\Delta\lambda/\lambda$) translates into a measurement of $\Delta\alpha/\alpha$ through the sensitivity parameter Q$\alpha$, an expression of the relative sensitivity of the transition frequency to variation in $\alpha$ (see Figure~\ref{fig: Qalpha} and \citealt{Berengut2013} for mathematical details). Similar shifts in the wavelengths of the molecular absorption lines of H$_2$ would indicate a variation in $\mu$.

The observations require far ultraviolet spectroscopy in the wavelength range of 1000--2000 Angstrom with a very high resolving power (R$\approx$200,000) to measure the potential wavelength differences. The success of the programme requires these observations for a sample spanning the full range of white dwarf gravities, with statistical completeness (20-30 objects) to determine whether any observed effect is real or to place scientifically useful limits on any variation.

\section{Physical Parameters}

\textbf{Overall Goal:} Determining basic parameters of a sample of white dwarfs (including temperature, $\log g$), as well as the chemical composition of the atmosphere, focusing on S, C, O, Mg, Fe and Si. Probe the dependence of fundamental physical constants on gravitational field for a subsample of white dwarfs at high spectral resolution, focusing on H$_2$, and high ionisation Fe and Ni (IV/V/VI) lines.

\subsection{Extrasolar planet composition \label{sec3:planet}}
In order to simulate what is needed to achieve the goals regarding extrasolar planet composition, here are some key assumptions, following the general method outlined in \citet{Buchan2024}: 
\begin{itemize}
    \item The distributions of white dwarf effective temperature, surface gravity and atmospheric type follow the Gaia white dwarf sample from \citet{Vincent2024}.
    \item Elements can be detected if their photospheric abundance surpasses a certain temperature-dependent threshold, estimated by interpolating white dwarf models (see Section~\ref{sec: observations}) linearly between 10,000 K and 20,000 K, and extrapolating beyond this range.
    \item The extrasolar material which pollutes white dwarfs initially form from a nebula whose composition matches nearby FGK stars \citep{Brewer2016}. 
    \item The mass of pollutants follows a collisional cascade distribution \citep{Dohnanyi1969}.
    \item Each accretion event is dominated by a single pollutant body, which accretes at a constant rate for a certain period of time. For H-dominated white dwarfs, we assume that any pollution is detected when accretion is on-going. For He-dominated white dwarfs, heavy elements can remain in the photosphere long after the end of accretion and so this assumption no longer holds. Therefore, we assume that accretion events last uniformly between 0 and 10 Myr, and they are observed at a time uniformly distributed between 0 and 20 Myr from the start of accretion
    \item The differential sinking of heavy elements in the white dwarf atmosphere follows \citet{Koester2020}, with convective overshoot included but no thermohaline mixing
\end{itemize}

\begin{figure*}[h]
    \centering
    \includegraphics[width=0.48\linewidth]{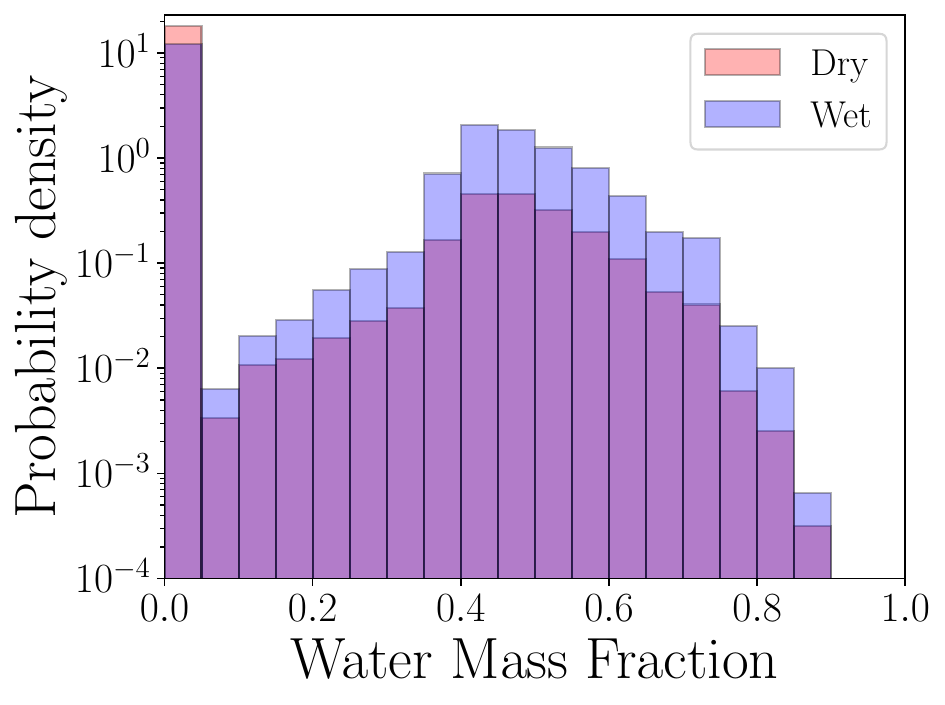}
    \includegraphics[width=0.48\linewidth]{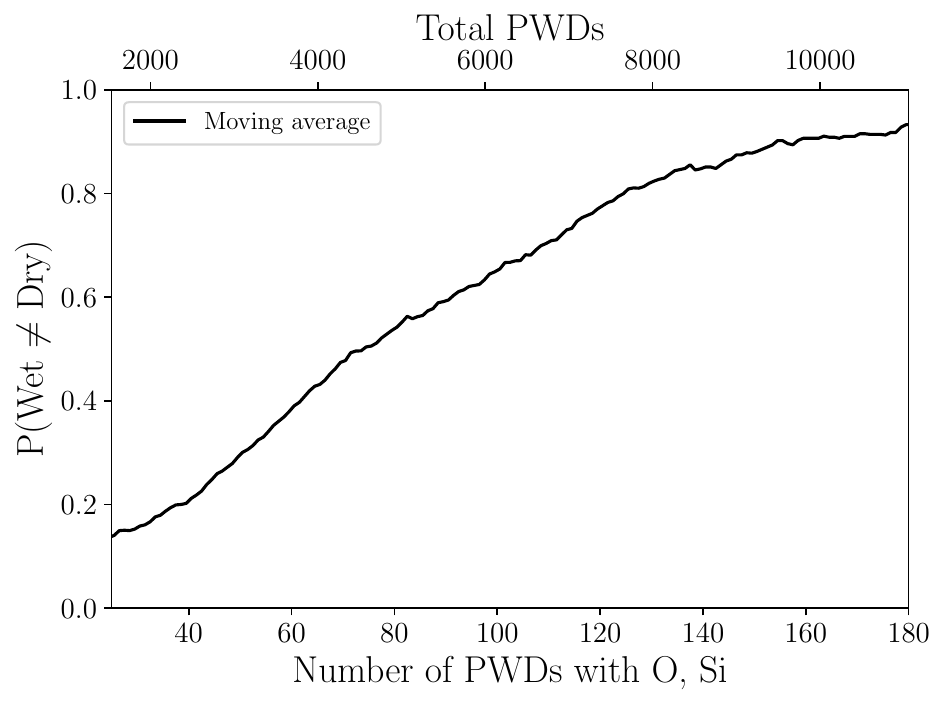}
    \caption{Left Panel: The distribution of water mass fraction for the `dry' and `wet' populations, to be tested with HWO observations. Right Panel: Number of polluted white dwarfs (PWDs) needed to distinguish between the two different models for the water mass fraction of the pollutants. A sample of 155 polluted white dwarfs with detections of O and Si is needed to distinguish the scenario at 90\% confidence.}
    \label{fig:water}
\end{figure*}

\textbf{Q1 How much water do extrasolar planets have?} We made the additional assumption that these pollutants can lose volatiles depending on their initial formation distance from their star (i.e., temperature), following Gibbs free energy minimisation (as in \citealt{Harrison2018}). The log (base 10) of the formation distance in AU is distributed between -1 and 1. This should be thought of as a proxy for formation temperature (between 80~K and 1680~K in our disc model), and hence the extent of volatile depletion. We explore two populations, as illustrated in Figure~\ref{fig:water}. For the `Wet' population, the formation distance distribution is biased towards high values, and for the `Dry' population it is biased towards low values. In order to distinguish between the wet and dry population to a 90\% confidence level, we require a sample of approximately 155 polluted white dwarfs with detections of Si and O within a precision of 0.1 dex (see Figure~\ref{fig:water}). This corresponds to identifying roughly 9500 polluted white dwarfs in total.

\textbf{Q2 What are the light elements in the cores of extrasolar planets?} Here, we focus on assessing the possibility of sulfur being the dominant light element in the core\footnote{We can design a similar experiment to assess the possibility of carbon being the dominant light element in the core.}, as opposed to a mixture of Si and O. In order to distinguish between a S-rich core and a Si/O-rich core, we made some additional assumptions:
\begin{itemize}
    \item The partitioning of elements between core and mantle is approximately Earth-like by default. For the `S rich core' and `Si/O rich core' cases, the resulting light element content in the core is stripped out and replaced with either S or a 50:50 O, Si mixture respectively. In all cases, the molar fraction of S in the mantle is not calculated within our model. Instead, it is randomly sampled from a distribution which is uniform between 0 and 0.05 by mole fraction.
    \item The distribution of core fractions (i.e., how much of each pollutant is core and how much is mantle) follows predictions of collisional evolution simulations as in \citet{Bonsor2020}.
\end{itemize}

\begin{figure}[h]
    \centering
    \includegraphics[width=\linewidth]{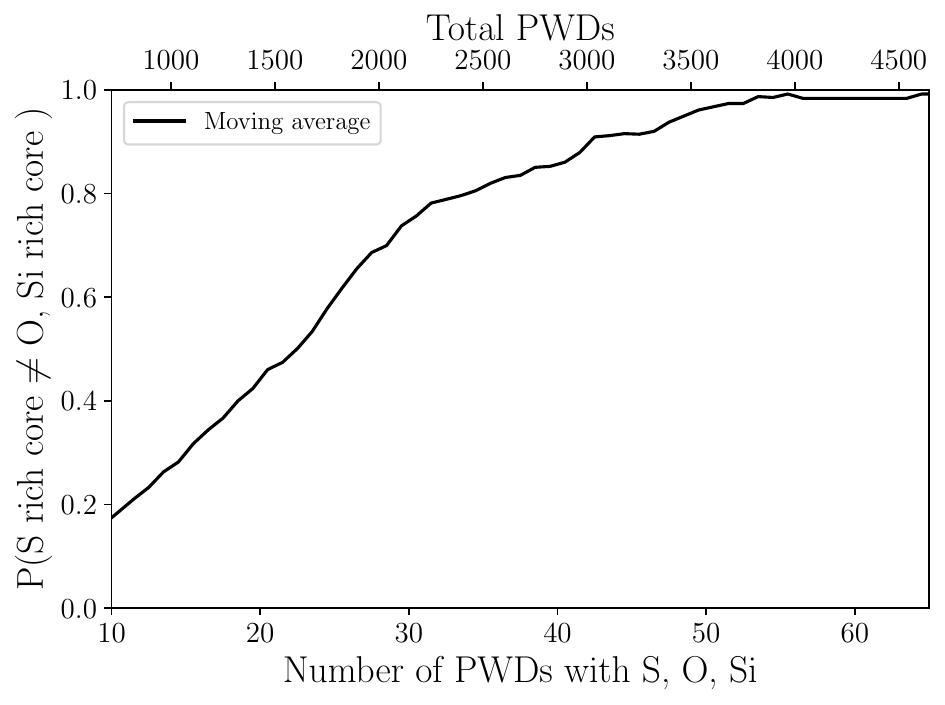}
    \caption{Similar to the right panel of Figure \ref{fig:water} except for distinguishing accretion of S-rich Fe cores and Si/O-rich Fe cores. A sample of 45 polluted white dwarfs with at least S, O, Si is needed to distinguish the scenario at 90\% confidence.}
    \label{fig:core}
\end{figure}

In order to achieve sufficient statistical power to distinguish between S-rich and Si/O-rich models with 90\% confidence, we require a sample of approximately 45 polluted white dwarfs with detections of S, O and Si with a precision of 0.1 dex (see Figure~\ref{fig:core}). This corresponds to identifying roughly 3200 polluted white dwarfs in total. A comparison with the current state of the art is summarized in Table~\ref{tab:number_PWD}.

\subsection{Fundamental physics}
Several theories of gravity allow the possibility of space and time variations in the low energy `constants' of Nature, either because of the presence of extra space dimensions or the non-uniqueness of the quantum vacuum state for the universe. Near massive gravitating bodies, different types of couplings between scalar fields and other fields can lead to an increase or decrease in the coupling constant strengths \citep{Magueijo2002}. In objects that are not too relativistic, such as white dwarfs, the total mass and total scalar charge are just proportional to the number of nucleons in the object. Compact objects with high mass and small radius, could exhibit detectable variations of fundamental constants \citep[e.g., ][]{FlambaumShuryak2008}.

To test the dependence of fundamental constant variations on gravity requires high resolution spectroscopic observations of high precision in wavelength to detect any difference between observed and predicted wavelengths of spectral lines. Furthermore, the measurements need to be made for a sample of white dwarfs that span the possible range of surface gravities (from log g = 7 -- 9) and temperatures, extending up to the hottest objects at $T_{\rm eff}$ $\approx$ 60,000 K. These objects will be a subset of the sample needed to study exoplanet composition, but the emphasis will be on observations to measure the small potential shifts in wavelength relative to laboratory values for high ionization species. The specific temperature is not important, but needs to be in a range that generates the high ionization absorption lines of interest. In the event that no shifts are detected, it is important that the measurements are precise enough to place scientifically valuable limits on the allowed variation, which can themselves rule out a number of gravitational models.

The basic measurement to be made for each line is the fractional wavelength shift for each line ($\Delta\lambda/\lambda$), which translates into a measurement of $\Delta\alpha/\alpha$ through the sensitivity parameter Q$_\alpha$, an expression of the relative sensitivity of the transition frequency to variation in $\alpha$ (see \citet{Berengut2013} for the mathematical details). To answer the question, `How do the fundamental constants $\alpha$ and $\mu$ vary in strong gravity?', a sample of bright white dwarfs with H$_2$ and/or Fe/Ni in their atmospheres is needed. As summarized in Table \ref{tab:wd_wv-measure-precision}, substantial progress requires a factor 4 larger sample than that existing coupled with a factor 4 improvement in the precision of wavelength measurement and, by implication, spectral resolution.

\begin{table*}[ht]
\caption{Number of white dwarfs required in the sample and wavelength measurement precision required for various levels of progress.}
    \label{tab:wd_wv-measure-precision}
\smallskip
\begin{center}
{\small
\begin{tabular}{lccc}  
\tableline
\noalign{\smallskip}
Physical Parameter & State of the Art & Substantial Progress & Major Progress\\
  & &(Enabling)& (Breakthrough)\\
\noalign{\smallskip}
\tableline
\noalign{\smallskip}
Number white dwarfs H$_2$ and high ionization Fe/Ni lines & 7 & 25 & 50 \\
        \noalign{\smallskip}
\tableline
\noalign{\smallskip}
        Precision of wavelength measurement (nm) & 0.0024 & 0.0006 & 0.0006\\
\noalign{\smallskip}
\tableline\
\end{tabular}
}
\end{center}
\end{table*}

\section{Description of Observations \label{sec: observations}}

\subsection{What data is needed?}

For the exoplanet composition objective, we need a sample of at least 45 heavily polluted white dwarfs with detections of S, C, O, Fe and Si. Given the current distribution of polluted white dwarfs, this requires a sample of 40,000 white dwarfs down to GALEX FUV magnitude of 20. A bright subset of these objects are expected to reveal H$_2$ and or high ionization Fe/Ni features, appropriate for studies of fundamental constant variation. A comparison with the current state of the art is shown in Table \ref{tab:observations_needed}.
\begin{table*}[ht]
\caption{The current state of the art is with HST/COS and HST/STIS. However, the resolving power is limited and it is only able to observe relatively bright white dwarfs. Therefore, there is only a small number of white dwarfs with detections of key elements and large uncertainties (see Tables~\ref{tab:number_PWD} and \ref{tab:wd_wv-measure-precision}).}
\smallskip
\begin{center}
{\small
\begin{tabular}{lcccc}  
\tableline
\noalign{\smallskip}
Observation requirement & State of the Art & Incremental Progress & Substantial Progress  & Major Progress \\
 & & &(Enabling)& (Breakthrough)\\
\noalign{\smallskip}
\tableline
\noalign{\smallskip}
Spectroscopy (exoplanet) & HST/COS, R$\approx$20,000 & R$\approx$20,000 & R$\approx$50,000 & R$\approx$60,000\\
        \noalign{\smallskip}
\tableline
\noalign{\smallskip}
        Spectroscopy (fund. constants) & HST/STIS,
R$\approx$50,000 & R$\approx$100,000 & R$\approx$200,000 & R$\approx$200,000\\
\noalign{\smallskip}
\tableline
\noalign{\smallskip}
        Wavelength Range (\r{A}) & $>1140$ & $>1140$ & $>1140$ & $>900$\\
        \noalign{\smallskip}
\tableline
\noalign{\smallskip}
        Amount of sky covered & \multicolumn{4}{c}{Single object}\\
        \noalign{\smallskip}
\tableline
\noalign{\smallskip}
        Mag. of target in GALEX FUV & $< 17$ mag & $< 18$ mag & $< 20$ mag & $< 20$ mag\\

\noalign{\smallskip}
\tableline\
\end{tabular}
}
\end{center}
\label{tab:observations_needed}
\end{table*}

\subsection{Detailed description of observations}
For both science cases, we need high-resolution ultraviolet spectra covering down to 900 Angstrom with a resolving power of at least 60,000.
\subsubsection{Spectral resolution}
\begin{figure*}[h!]
    \centering
    \includegraphics[width=0.4\linewidth,trim={0.5cm 0cm 0.8cm 1cm},clip]{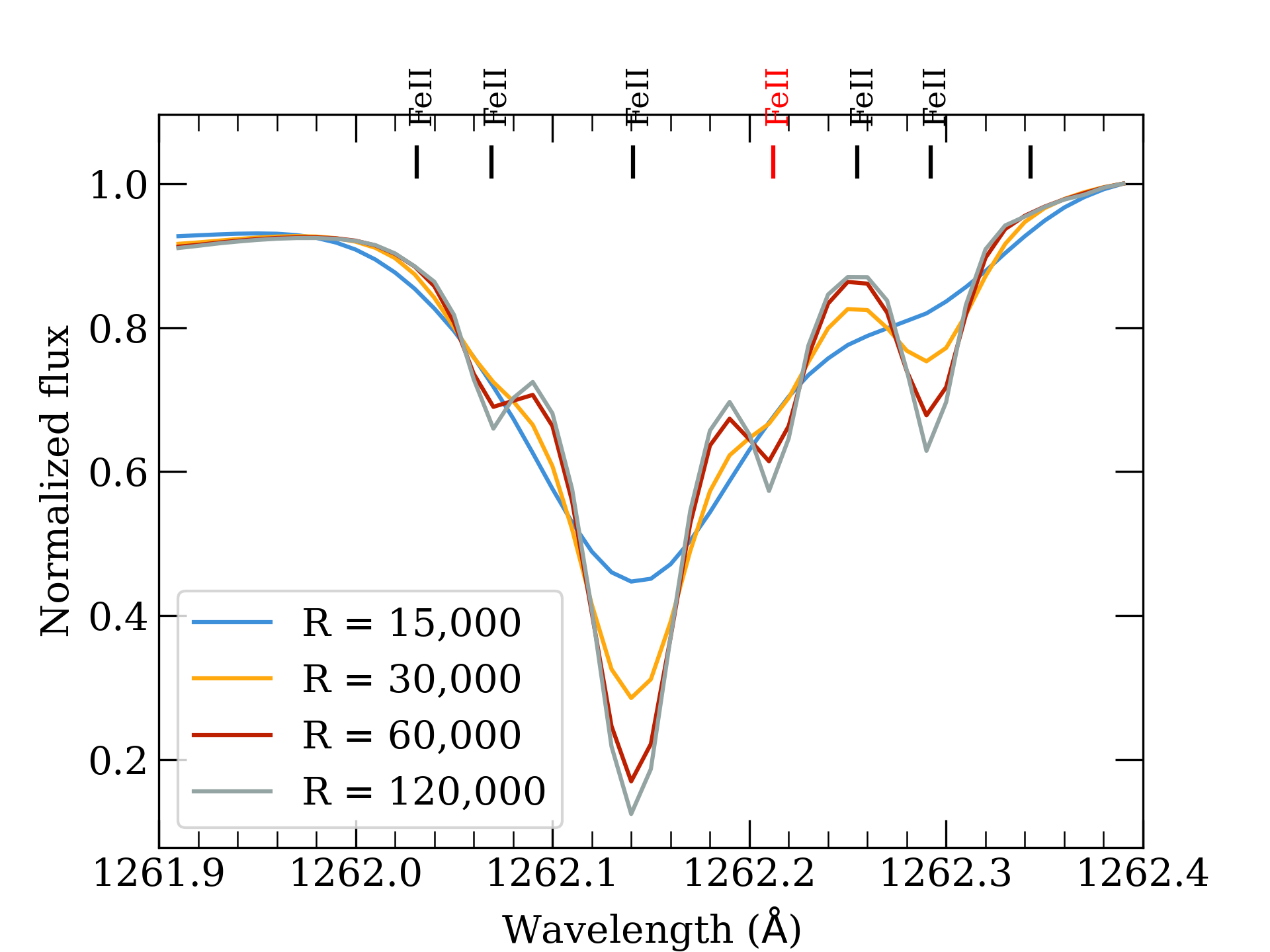}
\includegraphics[width=0.59\linewidth,trim={2.4cm 0cm 2.2cm 1.7cm},clip]{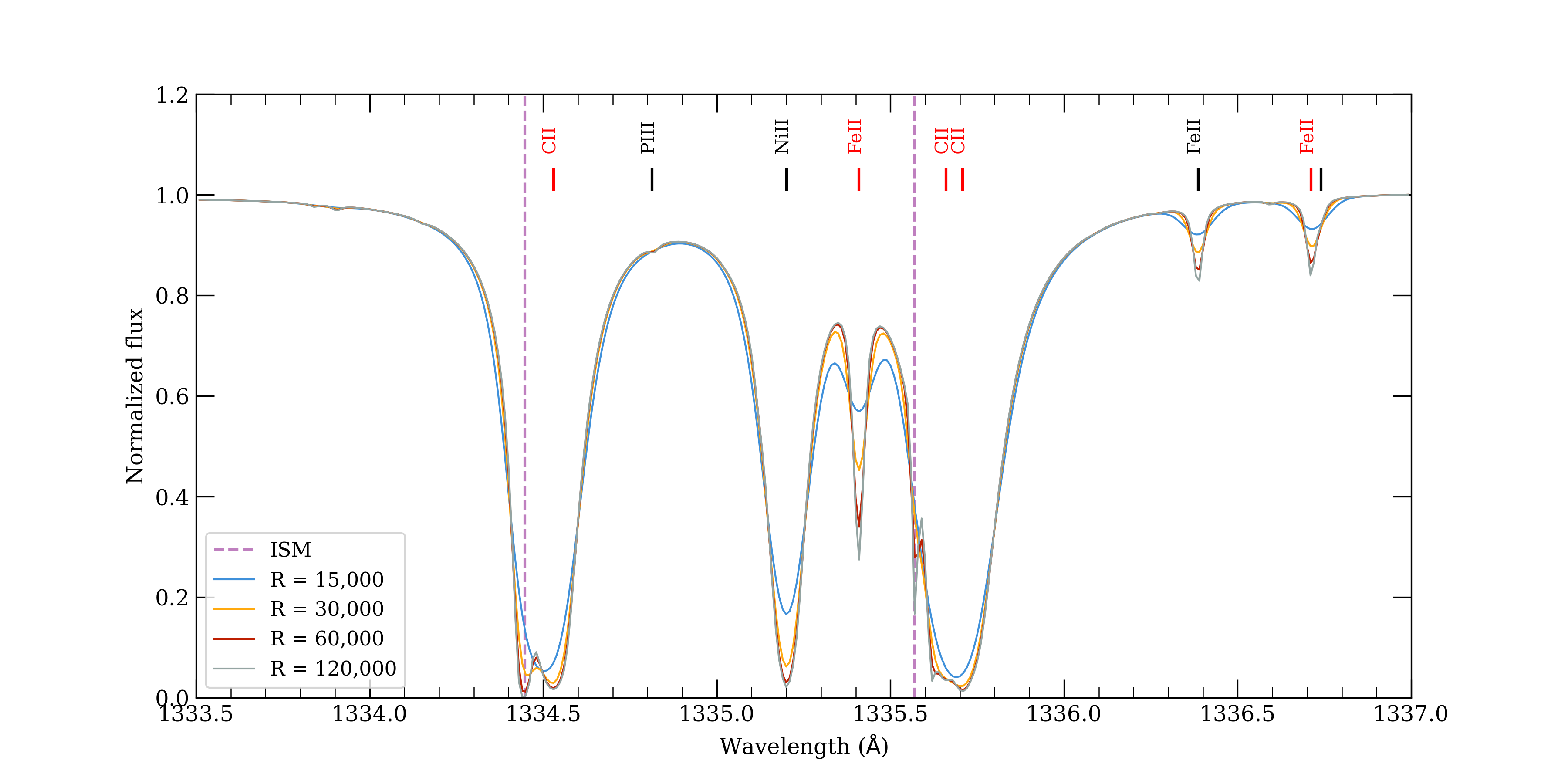}
    \caption{Simulated ultraviolet spectra for GD 40, a heavily polluted white dwarf from \citet{Jura2012}, assuming a resolving power of 15,000, 30,000, 60,000, and 120,000 for two wavelength regions. The left panel is around Fe II lines, where many weak lines are blended at a lower spectral resolution. The right panel is around the most important volatile element, C II, which tends to have complications from the interstellar medium. In order to properly determine the white dwarf photospheric abundances, a minimum resolving power of 60,000 is needed.}
    \label{fig:GD40_varR}
\end{figure*}

\begin{figure*}[h]
    \centering
    \includegraphics[width=0.49\linewidth]{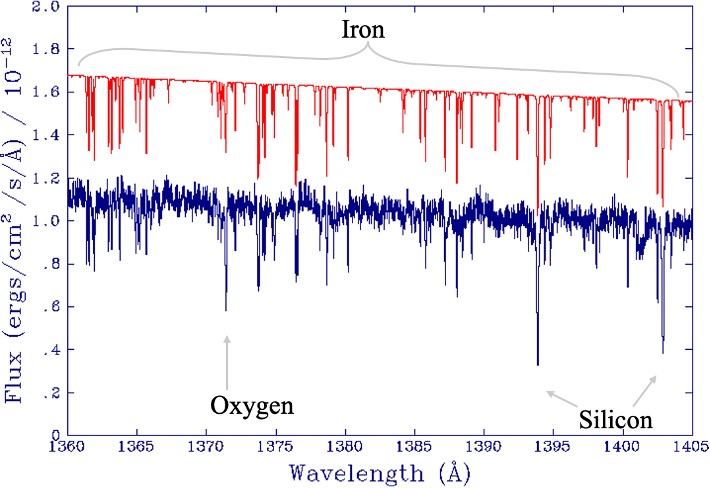}
     \includegraphics[width=0.49\linewidth]{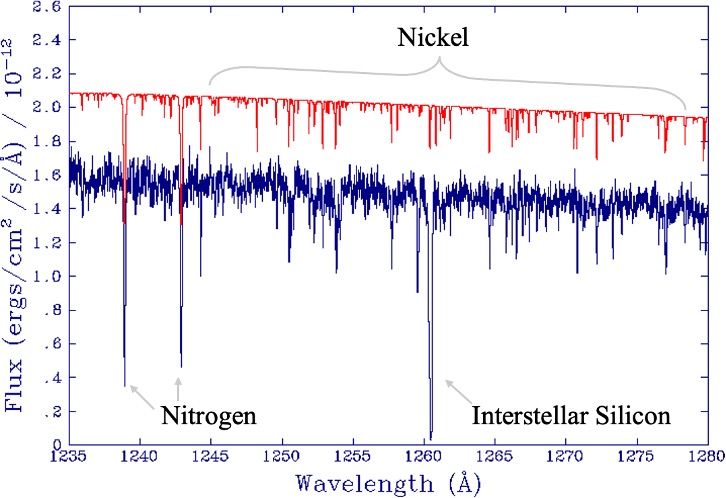}
    \caption{Section of an HST STIS spectrum of the hot white dwarf G191-B2B, recorded by the E140M (R$\approx$50,000) grating (blue), showing a large number of \ion{Fe}{5} and \ion{Fe}{6} lines in the left panel, and \ion{Ni}{5} and \ion{Ni}{6} lines in the right panel. The red line is a synthetic stellar spectrum computed for a model at the temperature and gravity of G191-B2B. It is offset from the observations for clarity.}
    \label{fig:G191-B2B}
\end{figure*}

As shown in Figure \ref{fig:GD40_varR}, in order to resolve the blended lines and separate the contribution from photospheric, interstellar, and circumstellar lines, a resolving power of at least 60,000 is needed. For measurement of the potential wavelength shifts from varying $\alpha$ and $\mu$, a higher precision is needed with a resolving power of 200,000. The current state of the art observations with STIS (resolving power of 50,000) is shown in Figure \ref{fig:G191-B2B}.

\subsubsection{UV cutoff}
\begin{figure*}[h]
    \centering
    \includegraphics[width=0.85\linewidth,trim={2.4cm 1.2cm 2.5cm 1.6cm},clip]{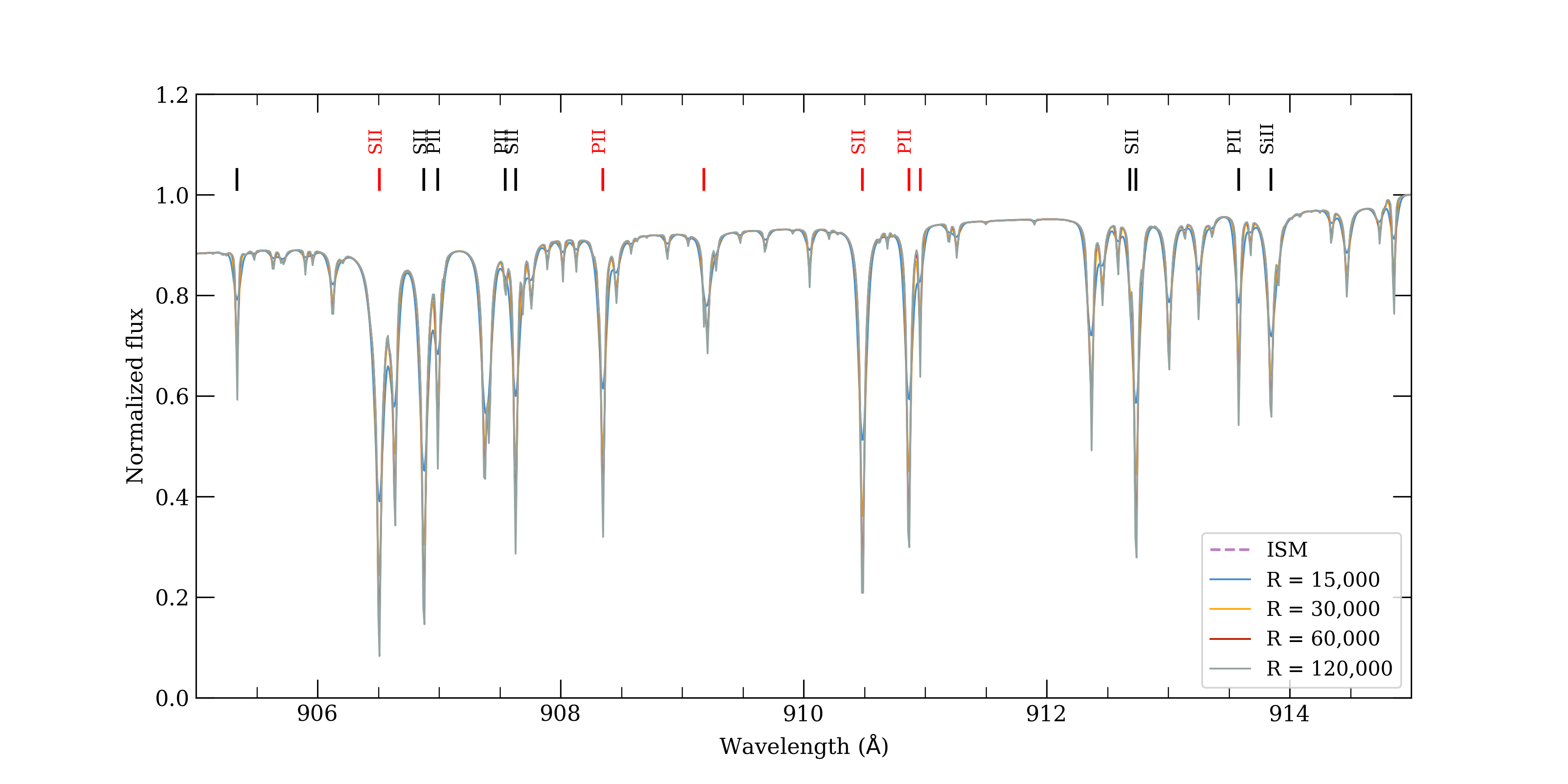}
    \includegraphics[width=0.85\linewidth,trim={2.4cm 1.2cm 2.5cm 1cm},clip]{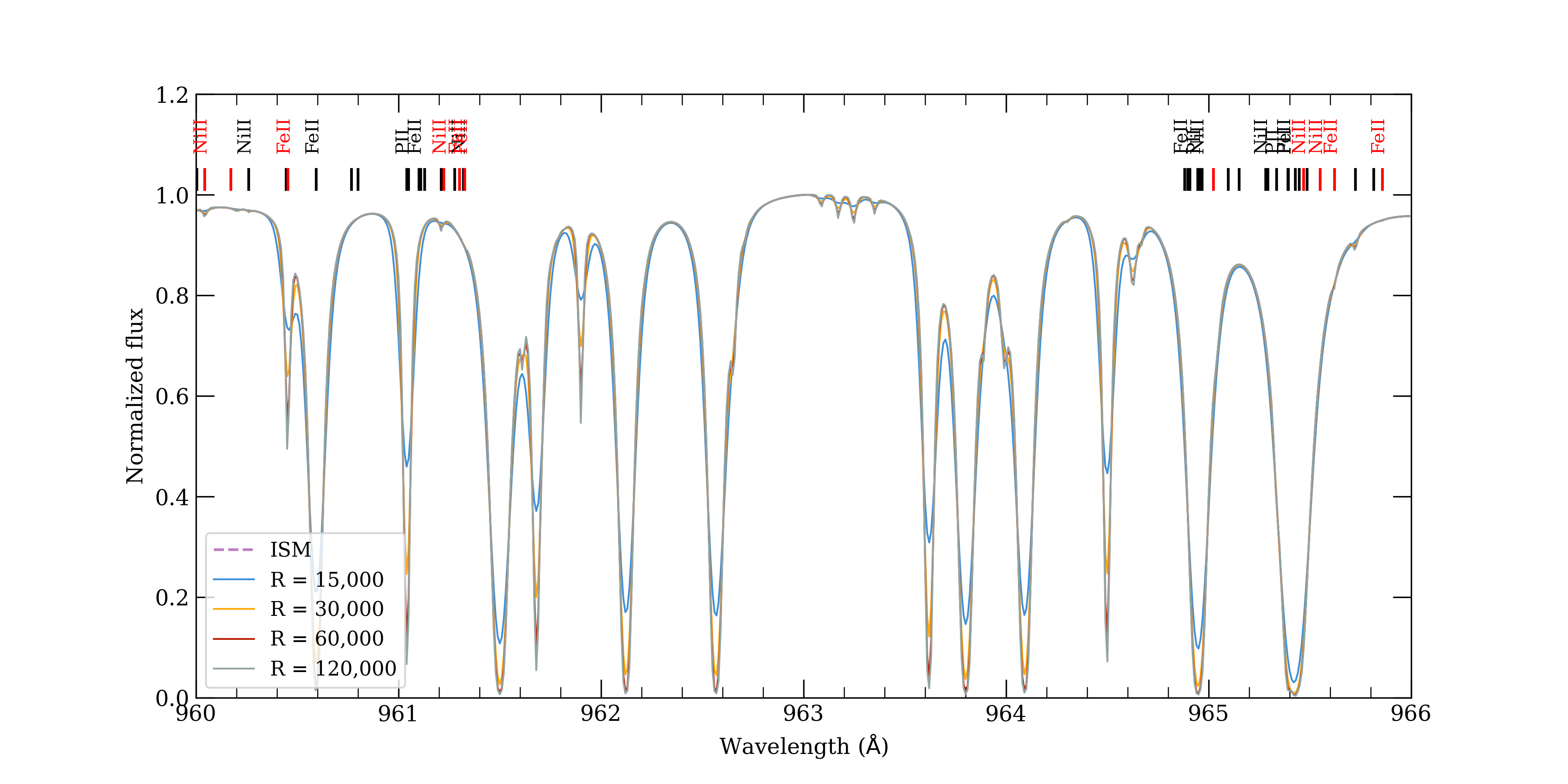}
    \includegraphics[width=0.85\linewidth,trim={2.4cm 0.5cm 2.5cm 1cm},clip]{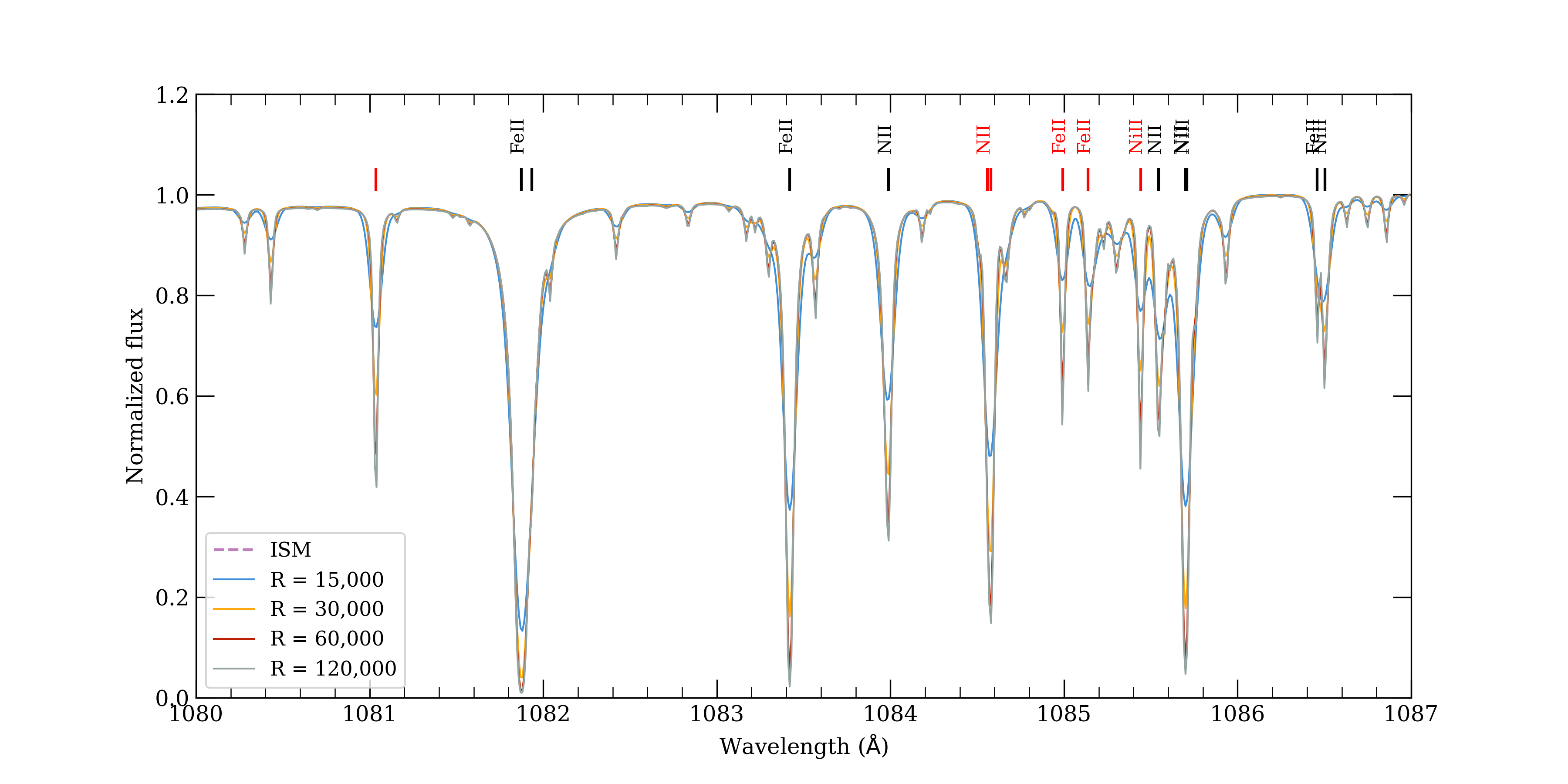}
    \caption{Example white dwarf model spectra highlight lines between 900 and 1100 Angstrom.}
    \label{fig:model-spec_1100A}
\end{figure*}

As shown in Table~\ref{tab:lines_interest}, we prefer a UV coverage down to at least 900 Angstrom in order to cover \ion{C}{2}, \ion{N}{2}, and \ion{S}{2} lines that are only accessible in the UV. Some example spectra are shown in Figure \ref{fig:model-spec_1100A}.

\begin{table*}[ht]
\caption{Tables in \LaTeXe}
\smallskip
\begin{center}
{\small
\begin{tabular}{ll}  
\tableline
\noalign{\smallskip}
Element & Strongest lines (\r{A}) \\
\noalign{\smallskip}
\tableline
\noalign{\smallskip}
 & \ion{C}{2} (ground state) 903.62, 903.96, 904.14, 904.48, 1036.34, 1037.02, 1334.53, 1335.66, 1335.71\\
        C & \ion{C}{1} (ground state) 945.56, 1277.29, 1277.56, 1328.83, 1329.01,1329.60\\
         & \ion{C}{2} 1009.86, 1010.09, 1010.38\\
         \noalign{\smallskip}
\tableline
          & \ion{N}{2} (ground state) 915.61, 915.96, 916.01, 916.70, 1083.99, 1084.58, 1085.55, 1085.70\\
         N & \ion{N}{1} (ground state) 963.99, 964.63, 965.04, 1134.16, 1134.41, 1134.98\\
           & \ion{N}{1} 1243.18, 1243.31\\
         \noalign{\smallskip}
\tableline
         O & \ion{O}{1} (ground state) 976.45, 977.96, 978.62, 1039.23, 1040.94, 1041.69\\
          & \ion{O}{1} 999.50, 1152.15, 1217.65\\
        \noalign{\smallskip}
\tableline
         & \ion{P}{2} (ground state) 961.04, 962.12, 962.57, 963.62, 963.80, 964.95, 965.33, 965.40, 966.51, \\
        P & 968.18, 969.36, 1154.00, 1159.09\\
        \noalign{\smallskip}
         & \ion{P}{2} 1015.45, 1064.78, 1249.83\\
         \noalign{\smallskip}
\tableline
         &  \ion{S}{2} (ground state) 906.51, 907.63, 910.49, 1250.58, 1253.81, 1259.52\\
        S & S I (ground state) 1425.03\\
         & \ion{S}{2} 1204.28, 996.01, 1000.49, 1006.09, 1006.26, 1014.44, 1019.52\\
        \noalign{\smallskip}
\tableline
        B & B I (ground state) 1362.46, 2496.78, 2497.73\\
        \noalign{\smallskip}
\tableline
        H$_2$ & Many H$_2$ transitions between 900 and 1600\\
        \noalign{\smallskip}
\tableline
        Fe & Thousands of Fe IV/V/VI transitions between 1000 and 1600\\
        \noalign{\smallskip}
\tableline
        Ni & Thousands of Ni IV/V/VI transitions between 1000 and 1600\\
\noalign{\smallskip}
\tableline\
\end{tabular}
}
\end{center}
\label{tab:lines_interest}
\end{table*}

\subsubsection{White Dwarf Modeling}
Detailed white dwarf atmospheric modeling is needed in order to back out the composition of the accreting exoplanetary material \citep[e.g.][]{Izquierdo2023}. The current abundance uncertainty is typically 0.2--0.3 dex, which is due to both the data quality and possibly some missing physics in white dwarf models. A better understanding of different systematics is crucial to achieve an abundance uncertainty of 0.1 dex. White dwarf atmosphere models are used in the fundamental physics work to facilitate the identification of spectral lines for the analyses.

\subsection{Why HWO?}

HWO is needed because volatile elements like C, S, and N, molecular hydrogen and high ionisation Fe and Ni can only be observed in the ultraviolet. The high-resolution spectrographs on the 30 m telescopes will only be able to detect O, Si, Fe and Mg.

\subsection{Other White Dwarf Science}

In this paper, we focus on the study of isolated non-pulsating white dwarfs. However, we note that there is significant and exciting science that HWO will be able to perform through studies of pulsating white dwarfs and white dwarfs in binary systems. Such studies could be usefully informed by production of additional Science Case Development Documents (SCDDs).

{\bf Acknowledgements.} S. Xu is supported by the international Gemini Observatory, a program of NSF NOIRLab, which is managed by the Association of Universities for Research in Astronomy (AURA) under a cooperative agreement with the U.S. National Science Foundation, on behalf of the Gemini partnership of Argentina, Brazil, Canada, Chile, the Republic of Korea, and the United States of America. MAB is grateful for support received from the Science and Technology Facilities Council UK (Grant numbers ST/Y003691/1 and UKRI1400). 

\bibliography{WD_Master.bib}

\end{document}